\documentclass{iau} 
\usepackage{graphicx}

\title[Optical spectroscopy of type-2 LINERs] %% give here short title %%
{Optical spectroscopy of type-2 LINERs}

\author[Laura Hermosa Mu{\~n}oz et al.]   %% give here short author list %%
{Laura Hermosa Mu{\~n}oz$^1$, 
 Sara Cazzoli$^1$, 
 Isabel M{\'a}rquez$^1$
 \and Josefa Masegosa$^1$}

\affiliation{$^1$Instituto de Astrof{\'i}sica de Andaluc{\'i}a - CSIC, Glorieta de la Astronom{\'i}a s/n 18008, Granada, Spain \\ email: {\tt lhermosa@iaa.es}}

\pubyear{2019}
\volume{356}  %% insert here IAU Symposium No.
\jname{Nuclear Activity in Galaxies Across Cosmic Time}
\editors{M. Povic, J. Masegosa, H. Netzer, P. Marziani, \\
P. Shastri, S. B. Tessema \& S. H. Negu, eds.}
\begin{document}

\maketitle

\begin{abstract}
Low-Ionisation Nuclear Emission-line Regions (LINERs) are the least luminous and the most numerous among the local population of Active Galactic Nuclei (AGN).
They can be classified as type-1 or type-2 if their optical spectra show or do not show, respectively, a broad component. It is associated with the presence of a Broad Line Region (BLR) in these systems. However, recent studies have proven that the classification of type-1 LINERs may be controversial, since space- and ground-based spectroscopy provide contradicting results on the presence of very broad components (\cite[Cazzoli et al. 2018]{Cazzoli18}).
We have studied the nuclear spectra of 9 type-2 LINERs with intermediate spectral resolution HST/STIS data. We present the results on our analysis of the different spectral components, and discuss the eventual presence of BLR components in type-2 LINER galaxies, together with the possible presence of outflows, both in comparison with type-1 LINERs. We have found a BLR component in 7 out of the 9 analysed objects within the HST/STIS data.
%% add here a maximum of 10 keywords, to be taken form the file <Keywords.txt>
\end{abstract}

\firstsection % if your document starts with a section,
              % remove some space above using this command.
\section{Introduction}

LINERs are low luminosity AGNs whose spectra are dominated by low-ionisation emission lines (\cite[Heckman 1980]{Heckman1980}). Their true nature is still uncertain (\cite[Ho 2008]{Ho2008}, \cite[M{\'a}rquez et al. 2017]{Marquez2017}) as their spectral features can be explained with other models apart from an AGN (see e.g. for shock ionisation models: \cite[Heckman 1980]{Heckman1980}, \cite[Dopita, \& Sutherland 1995]{Dopita1995}; and for photoionisation by post-AGB stars: \cite[Binette et al. 1994]{Binette1994}, \cite[Papaderos et al. 2013]{Papaderos2013}). As other AGNs, they are optically classified depending on whereas the BLR is accessible along the line of sight, and hence a very broad Balmer line is detected.
Type-2 LINERs have been classified by \cite[Ho et al. (1993)]{Ho1993} as not showing a broad component in the H$\alpha$ line, although it is visible in type-1 AGNs, as we have a direct view of the innermost parts of the AGN. However, it has been reported that the broad component is not visible in some optically classified type-1 LINERs, with differing results from space- to ground-based data (\cite[Cazzoli et al. 2018]{Cazzoli18}). 
Additionally, the emission line profiles may be affected by the presence of outflows. These components, whose presence could compromise the detection of weak BLR components in LINERs, can be detected in [S\,II]$\lambda \lambda$6716,6731\AA\., H$\alpha \lambda$6564\AA\., [N\,II]$\lambda \lambda$6548,6584\AA\, and [O\,I]$\lambda \lambda$6300,6363\AA\ lines. 

We have analysed archival data of the 9 type-2 LINERs from the sample of 82 LINERs by \cite[Gonz{\'a}lez-Mart{\'i}n et al. (2009)]{OGM09} with Hubble Space Telescope (HST)/Space Telescope Imaging Spectrograph (STIS) nuclear spectra. We have completed the sample with ground-based spectra from the Double Spectrograph / Palomar (\cite[Ho et al. 1993]{Ho1993}; \cite[Ho et al. 1995]{Ho1995}) for 8 out of 9 LINERs (NGC\,4676B was not studied in the Palomar sample). 
The main objective is to study the presence of different kinematic components in the nuclei of these galaxies to unveil the true nature of type-2 LINERs.

\section{Methodology}

The stellar subtraction and spectral fitting were done as in \cite[Cazzoli et al. (2018)]{Cazzoli2018} (hereafter C18).
Spectra from both space- and ground-based data were retrieved from the archival already fully reduced. The gratings for both data-sets covered the wavelength range where [S\,II], H$\alpha$-[N\,II] and, in some cases, [O\,I] emission lines are available.

As the host galaxy contribution on the nuclear spectra of a low-luminosity AGN could be significant, the starlight should be subtracted before fitting the lines. 
However, for the HST/STIS data the wavelength range is small (572 \AA) for a proper stellar continuum modelling. Nevertheless, thanks to the small aperture of the instrument, \cite[Constantin et al. (2015)]{Constantin2015} proved that the correction is negligible for HST/STIS spectra.
Thus the starlight was modelled and subtracted only for the Palomar data using a penalised PiXel fitting analysis (pPXF version 4.71
\cite[Cappellari, \& Emsellem 2004]{Cappellari2004}; \cite[Cappellari 2017]{Cappellari2017}).  

The emission lines were modelled applying a non-linear least-squares minimization and curve-fitting routine (\textsc{lmfit}) implemented in Python. We used a maximum of two Gaussians per forbidden line and narrow H$\alpha$ plus a broad Gaussian for H$\alpha$ when needed. To prevent overfitting, we used the standard deviation of a region of the continuum without absorption or emission lines ($\varepsilon_{c}$). This value was compared to the standard deviation of the residuals under the lines after adding the Gaussian component. If it was higher than 3$\varepsilon_{c}$ a new component was added. 
All the lines were tight to have the same velocity (v) and velocity dispersion ($\sigma$) as the [S\,II] or [O\,I] lines. We used them as reference for H$\alpha$-[N\,II] as they are usually unblended in the spectra, thus causing less uncertainties when modelling the line profiles (see C18 for more details).  

\section{Results}

\begin{figure}
 \begin{center}
 \includegraphics[clip,trim=0.5cm 11.8cm 1.4cm 9cm,width=\textwidth]{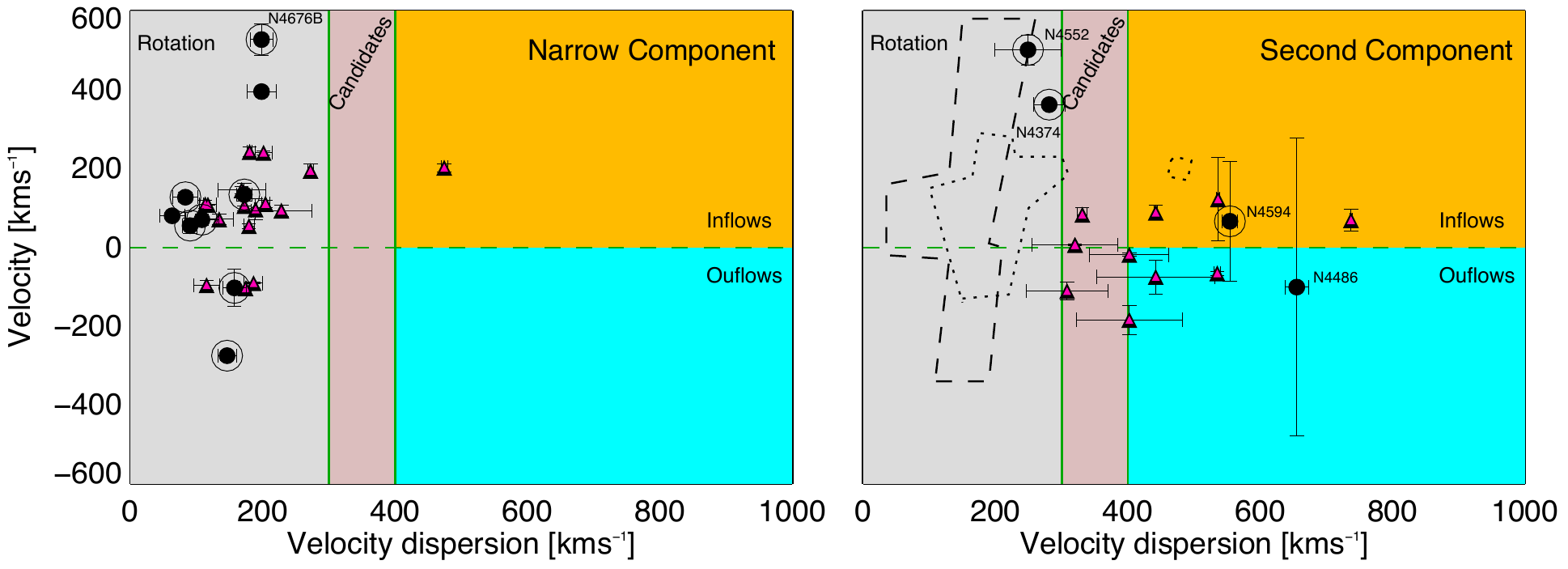} 
 \caption{Kinematic classification of the components from the spectral modelling. The velocity and velocity dispersion for the narrow (left) and secondary (right) components are shown for both type-2 (black circles) and type-1 LINERs (red triangles from C18). Grey bands indicate rotational motions; pink regions are candidates to non rotational motions; yellow and blue zones indicate inflows and outflows, respectively. Black circles surround galaxies where a broad component was detected. The dashed (dotted) line indicates the region occupied by the narrow component in type-2 (type-1) LINERs.}
  \label{fig1}
  \end{center}
\end{figure}

The kinematics (v and $\sigma$) of each Gaussian component used in the fit (except for the broad component originated in the BLR) can be associated to either rotational or non-rotational motions that occur in the host galaxy.
The main result of this analysis is summarised in Fig.~\ref{fig1}. The narrowest Gaussian profile is called here \textit{narrow component}, and a broader Gaussian which is needed in all emission lines for some LINERs is called \textit{secondary component}. The narrow component is generally associated to rotation coming from the galactic disk, whereas the secondary component could be due to rotational motions or non-rotational motions (outflows or inflows). Dividing lines in Fig.~\ref{fig1} have been established by C18 measuring the gas velocity field from the 2D spectra of their sample of type-1.9 LINERs, and estimating the maximum broadening of the emission lines associated to rotational motions. Their conservative upper limit for the rotational component is $\sim$ 400 km$s^{-1}$, although the values of $\sigma$ in their sample are well below this limits (see Sect.\,5.2 in C18).

For the type-2 LINERs within our sample, all the narrow components found in the modelling are consistent with being produced by rotational motions, as initially expected. 
Only four galaxies needed a secondary component to reproduce the observed line profiles. We have found that this component is associated to both rotational (NGC\,4552 \& NGC\,4374) and non-rotational motions (NGC\,4594 \& NGC\,4486). Thus the latter objects are candidates to have outflows.
As for the BLR component, we found that it is needed to reproduce the H$\alpha$ profile in 7 out of the 9 galaxies from the HST/STIS data and in 2 out of 8 in the Palomar data. Three of the HST detections (NGC\,3245, NGC\,4594 and NGC\,4736) were already reported by \cite[Constantin et al. (2015)]{Constantin2015}. Our results for the Full Width at Half Maximum and the flux contribution of the BLR component agree with theirs. 
Therefore, the true nature of most LINERs needs to be revisited with high spectral resolution.

\section{Conclusions}

The main conclusions of the work \cite[Hermosa-Mu{\~n}oz et al. (2020)]{HM2020} are summarised below:

\begin{itemize}
    \item A BLR component is detected in 7 out of the 9 objects in the HST/STIS data, with 3 already reported in \cite[Constantin et al. (2015)]{Constantin2015}. 
    \item Two out of the 8 objects from the Palomar sample show a broad component that was not found by \cite[Ho et al.(1995)]{Ho1995}.
    \item The detection of a broad component is favoured in the HST data as the spatial resolution is 10 times better than for the Palomar data.
    \item NGC\,4594 is the only object with a broad detection in both ground- and space-based spectra. We propose its reclassification as a type-1 LINER.
    \item A secondary component consistent with outflows is detected in NGC\,4594 and NGC\,4486.
\end{itemize}

\end{document}